\documentclass[12pt]{article}
\usepackage{epsfig,cite,amsmath}

\def\beq{\begin{equation}}
\def\eeq{\end{equation}}
\def\beqar{\begin{eqnarray}}
\def\eeqar{\end{eqnarray}}
\def\barr#1{\begin{array}{#1}}
\def\earr{\end{array}}
\def\bfi{\begin{figure}}
\def\efi{\end{figure}}
\def\btab{\begin{table}}
\def\etab{\end{table}}
\def\bce{\begin{center}}
\def\ece{\end{center}}

\def\text{\textstyle}

\def\be{\beta}

\def\reffi#1{\mbox{Fig.~\ref{#1}}}

\def\citere#1{\mbox{Ref.~\cite{#1}}}
\def\citeres#1{\mbox{Refs.~\cite{#1}}}

\def\mathswitchr#1{\relax\ifmmode{\mathrm{#1}}\else$\mathrm{#1}$\fi}

\newcommand{\PA}{\mathswitchr A}

\newcommand{\Pt}{\mathswitchr t}

\def\mathswitch#1{\relax\ifmmode#1\else$#1$\fi}

\newcommand{\Mt}{\mathswitch {m_\Pt}}

\newcommand{\MA}{\mathswitch {M_\PA}}

\newcommand{\mt}{\Mt}

\newcommand{\Mgl}{M_{\tilde{\mathrm{g}}}}

\newcommand{\tsf}{\theta\kern-.20em_{\tilde{f}}}
\newcommand{\tsfp}{\theta\kern-.20em_{\tilde{f}\prime}}
\newcommand{\tsq}{\theta\kern-.15em_{\tilde{q}}}

\newcommand{\lsim}
{\;\raisebox{-.3em}{$\stackrel{\displaystyle <}{\sim}$}\;}
\newcommand{\gsim}
{\;\raisebox{-.3em}{$\stackrel{\displaystyle >}{\sim}$}\;}

\newcommand{\cp}{{\cal CP}}

\newcommand{\VL}{\left( \begin{array}{c}}
\newcommand{\VR}{\end{array} \right)}
\newcommand{\ML}{\left( \begin{array}{cc}}
\newcommand{\MLd}{\left( \begin{array}{ccc}}
\newcommand{\MLv}{\left( \begin{array}{cccc}}
\newcommand{\MR}{\end{array} \right)}

\newcommand{\BC}{\begin{center}}
\newcommand{\EC}{\end{center}}
\newcommand{\BE}{\begin{equation}}
\newcommand{\EE}{\end{equation}}
\newcommand{\BEA}{\begin{eqnarray}}
\newcommand{\BEAnn}{\begin{eqnarray*}}
\newcommand{\EEA}{\end{eqnarray}}
\newcommand{\EEAnn}{\end{eqnarray*}}

\newcommand{\id}{{\rm 1\kern-.12em
\rule{0.3pt}{1.5ex}\raisebox{0.0ex}{\rule{0.1em}{0.3pt}}}}

\def\draftdate{\relax}
\def\mda{\relax}
\def\mua{\relax}
\def\mla{\relax}
\def\draft{
\def\thtystars{******************************}
\def\sixtystars{\thtystars\thtystars}
\typeout{}
\typeout{\sixtystars**}
\typeout{* Draft mode!
         For final version remove \protect\draft\space in source file
*}
\typeout{\sixtystars**}
\typeout{}
\def\draftdate{\today}
\def\mua{\marginpar[\boldmath\hfil$\uparrow$]%
                   {\boldmath$\uparrow$\hfil}%
                    \typeout{marginpar: $\uparrow$}\ignorespaces}
\def\mda{\marginpar[\boldmath\hfil$\downarrow$]%
                   {\boldmath$\downarrow$\hfil}%
                    \typeout{marginpar: $\downarrow$}\ignorespaces}
\def\mla{\marginpar[\boldmath\hfil$\rightarrow$]%
                   {\boldmath$\leftarrow $\hfil}%
                    \typeout{marginpar:
$\leftrightarrow$}\ignorespaces}
\def\Mua{\marginpar[\boldmath\hfil$\Uparrow$]%
                   {\boldmath$\Uparrow$\hfil}%
                    \typeout{marginpar: $\Uparrow$}\ignorespaces}
\def\Mda{\marginpar[\boldmath\hfil$\Downarrow$]%
                   {\boldmath$\Downarrow$\hfil}%
                    \typeout{marginpar: $\Downarrow$}\ignorespaces}
\def\Mla{\marginpar[\boldmath\hfil$\Rightarrow$]%
                   {\boldmath$\Leftarrow $\hfil}%
                    \typeout{marginpar:
$\Leftrightarrow$}\ignorespaces}
\overfullrule 5pt
\oddsidemargin -15mm
\marginparwidth 29mm
}

\begin{document}
\thispagestyle{empty}

\def\thefootnote{\fnsymbol{footnote}}

\begin{flushright}
DCPT/02/152\\
IPPP/02/76\\
hep-ph/0301111 \\
\end{flushright}

\mbox{}

\vspace{2cm}

\begin{center}

{\large\sc {\bf The LHC / LC Study Group}}

\vspace{0.4cm}

{\large\sc {\bf and the Snowmass Points and Slopes}}%
\footnote{Talk given by G.~Weiglein at SUSY02, June 2002, DESY, Germany}

\vspace{1cm}

{\sc 
G.~Weiglein%
\footnote{email: Georg.Weiglein@durham.ac.uk}
}

\vspace*{1cm}

{\sl
Institute for Particle Physics Phenomenology,\\ 
University of Durham, Durham DH1 3LE, U.K.
}

\end{center}

\vspace*{1.2cm}

\begin{abstract}
The ``LHC / LC Study Group'' investigates how analyses at the LHC could
profit from results obtained at a future Linear Collider and vice versa.
Some of the activities of this recently formed working group 
are briefly summarised. 
The LHC / LC Study Group home page is
{\tt www.ippp.dur.ac.uk/$\sim$georg/lhclc}.
The ``Snowmass Points and Slopes'' (SPS) are a set of benchmark
points and parameter lines in the MSSM parameter space
corresponding to different scenarios in the search
for Supersymmetry at present and future experiments. This set of
benchmarks was agreed upon at the 2001 ``Snowmass Workshop on the Future
of Particle Physics'' as a consensus based on different existing proposals.
Further information about the SPS can be found under
{\tt www.ippp.dur.ac.uk/$\sim$georg/sps}.
\end{abstract}

\def\thefootnote{\arabic{footnote}}
\setcounter{page}{0}
\setcounter{footnote}{0}

\newpage


\section{The LHC / LC Study Group}

The aim of the LHC / LC Study Group is to investigate a possible
cross-talk between the LHC and a future Linear Collider (LC) and to
study in how far analyses carried out at one of the machines could profit 
from results obtained at the other machine. Mutual benefits could occur 
both at the level of a combined interpretation of Hadron Collider and 
Linear Collider data and at the level of combined analyses of the data,
where results obtained at one machine
could directly influence the way analyses are carried out the other
machine.

Topics under study comprise the physics of weak and strong
electroweak symmetry breaking, Supersymmetric models, new gauge
theories, models with extra dimensions, and electroweak and QCD
precision physics. For these studies it is assumed that the LC comes
into operation about half a decade after the start of the LHC. During
simultaneous running of both machines there is obviously the highest
flexibility for adapting analyses carried out at one machine according
to the results obtained at the other machine. The LC results could in
this context also serve as an input for a second phase of LHC running
concerning different possible upgrade options. 

\begin{figure}[ht]
\bce
\includegraphics[width=16cm]{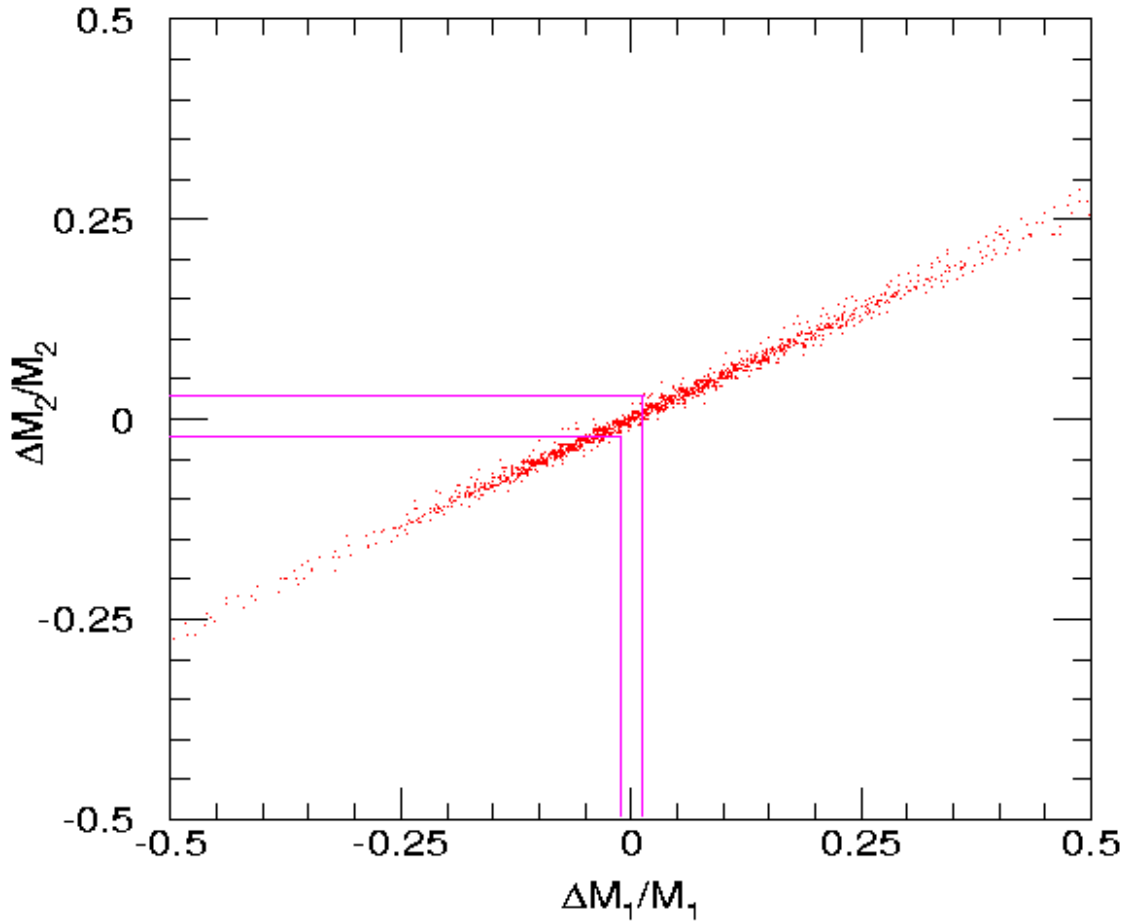}
\ece
\caption{The relative accuracy in the determination of the masses of the
lightest and the next-to-lightest neutralino at the LHC, taken from
\citere{atlastdr}. The narrow band overlayed in the plot shows the
improvement in the mass determination of the next-to-lightest neutralino
at the LHC if the measurement of the LSP mass at the LC is used as
input.
}
\label{fig:lhclc}
\end{figure}

To mention just one example of studies carried out in the working group, 
the determination of masses of
Supersymmetric particles at the LHC could profit from the precise measurement 
of the lightest Supersymmetric particle (LSP) at the LC. 
As a trivial illustration of the possible benefits of a precise
measurement of the LSP mass at the LC, \reffi{fig:lhclc} shows the
relative accuracy of the determination of the mass of the 
next-to-lightest neutralino at the LHC, which is highly correlated with 
the accuracy of the LSP mass determination~\cite{atlastdr}. 
The prospective accuracy of 
the LSP mass measurement at the LC is indicated by a narrow band, which
obviously leads to a drastic improvement of the mass determination of
the next-to-lightest neutralino at the LHC.

The LHC / LC Study Group has been forming over the last months and is still
further expanding. Currently there are about 140 working group members
(members of ATLAS, CMS, LC working groups, a Tevatron contact person,
and theorists). Further information about the working group can be
obtained from {\tt rohini.godbole@cern.ch}, {\tt paige@bnl.gov}, 
{\tt Georg.Weiglein@durham.ac.uk} or from the LHC / LC Study Group web 
page {\tt www.ippp.dur.ac.uk/$\sim$georg/lhclc}. 


\section{The Snowmass Points and Slopes (SPS)}

In the unconstrained version of the Minimal Supersymmetric extension of
the Standard Model (MSSM) no particular Supersymmetry (SUSY) breaking 
mechanism is assumed, but rather a parameterisation of all possible soft 
SUSY breaking terms is used.
This leads to more than a hundred parameters
(masses, mixing angles, phases) in this model in addition to the ones of
the Standard Model. For performing detailed simulations of experimental
signatures within detectors of high-energy physics experiments it is
clearly not practicable to scan over a multi-dimensional parameter space. 
One thus often concentrates on certain ``typical'' benchmark scenarios.

The Snowmass Points and Slopes (SPS)~\cite{sps} are a set of benchmark
scenarios in the MSSM parameter space which was agreed upon at
the 2001 ``Snowmass Workshop on the Future
of Particle Physics'' as a consensus based on different existing
proposals~\cite{existprop}.
The SPS consist of model
lines (``slopes''), i.e.\ continuous sets of parameters depending on one 
dimensionful parameter and specific benchmark points, where
each model line goes through one of the benchmark points. The SPS should
be regarded as a recommendation for future studies of SUSY phenomenology, 
but of course are not meant as an exclusive and for all purposes sufficient
collection of SUSY models. They mainly focus on ``typical'' scenarios
within the three currently most prominent SUSY-breaking mechanisms,
i.e.\ minimal supergravity (mSUGRA)~\cite{msugra}, gauge-mediated SUSY
breaking (GMSB)~\cite{gmsb}, and anomaly-mediated
SUSY breaking (AMSB)~\cite{amsb}.
Furthermore they contain examples of 
``more extreme'' scenarios, e.g.\ a ``focus point''
scenario~\cite{focuspoint} with a rather heavy SUSY spectrum, indicating
in this way different possibilities for SUSY phenomenology that can 
be realised within the most commonly used SUSY breaking scenarios.

The SPS comprise ten benchmark points, from which six correspond to an 
mSUGRA scenario, one is an mSUGRA-like scenario with non-unified gaugino
masses, two refer to the GMSB scenario, and one to the AMSB scenario. 
Seven of these benchmark points are attached to model lines, while the
remaining three are supplied as isolated points.
In studying the benchmark scenarios the model lines should prove
useful in performing more general analyses of typical SUSY signatures, 
while the specific points indicated on the lines are proposed to be 
chosen as the first sample points for very detailed (and thus 
time-consuming) analyses. The concept of a model line means of course
that more than just one point should be studied on each line. 
Results along the model lines can often then be roughly
estimated by interpolation.

The SUSY-breaking scenarios mentioned above are characterised by a few
input parameters (in the mSUGRA scenario, for instance, these are the
scalar mass parameter $m_0$, the gaugino mass parameter
$m_{1/2}$, the trilinear coupling $A_0$, the ratio of the Higgs vacuum
expectation values, $\tan\beta$, and the sign of the Supersymmetric
Higgs mass parameter, $\mu$). The mass spectra of the
SUSY particles in the mSUGRA, GMSB and AMSB scenarios are obtained via 
renormalisation group running from the scale of the high-energy parameters 
of the SUSY-breaking scenario to the weak scale.
The low-energy parameters obtained in this way are then used as input
for calculating the predictions for the production cross sections and 
for the decay branching ratios of the SUSY particles.

An important aspect in the philosophy behind the benchmark scenarios is 
that the low-energy MSSM parameters are defined to be the actual
benchmark rather than the high-energy input parameters $m_0$, $m_{1/2}$,
etc. A specification of the benchmark scenarios in terms of the latter
parameters is merely understood as an abbreviation for the low-energy
phenomenology (it also depends on the particular program used for
relating the high-energy input parameters to the low-energy MSSM
parameters).

While certain sets of low-energy MSSM parameters have been fixed as 
benchmarks in the SPS by definition (which in principle could have been done
without resorting at all to scenarios like mSUGRA, GMSB and AMSB),
the evaluation of the mass spectra and decay branching ratios from the
MSSM benchmark parameters should be carried out with the tools and at the
level of sophistication being most appropriate for the particular
application one is interested in. If detailed comparisons between 
different experiments or different colliders are carried out, it would 
clearly be advantageous to use the same results for the mass spectra and 
the branching ratios.

The main qualitative difference between the SPS (and also the recent
proposals for post-LEP benchmarks in \citere{existprop}) 
and the benchmarks used previously for
investigating SUSY searches at the LHC, the Tevatron and a future Linear
Collider is that scenarios with small
values of $\tan\be$, i.e.\ $\tan\be \lsim 3$, are disfavoured as a result
of the Higgs exclusion bounds obtained at LEP. Consequently, there is
more focus now on scenarios with larger values of $\tan\be$ than in
previous studies. Concerning the SUSY phenomenology, intermediate and
large values of $\tan\be$, $\tan\be \gsim 5$,
have the important consequence that there is
in general a non-negligible mixing between the two staus (and an even
more pronounced mixing in the sbottom sector), leading to a significant mass
splitting between the two staus so that the lighter stau becomes the
lightest slepton. Neutralinos and charginos therefore decay
predominantly into staus and taus, which is experimentally more
challenging than the dilepton signal resulting for instance from the decay 
of the second lightest neutralino into the lightest neutralino and a
pair of leptons of the first or the second generation.

Large values of $\tan\be$ can furthermore have important consequences
for the phenomenology in the Higgs sector, as the couplings of the heavy
Higgs bosons $H$, $A$ to down-type fermions are in general enhanced. 
For sizable values of $\mu$ and $m_{\tilde g}$ the $hb\bar b$ coupling
receives large radiative corrections from gluino loop corrections, 
which in particular affect the branching ratio BR($h \to \tau^+\tau^-$).

The main features of the SPS benchmarks are listed in the following:

\begin{description}

\item[SPS 1: ``typical'' mSUGRA scenario] 

\mbox{}

This scenario consists of a ``typical'' mSUGRA point with an
intermediate value of 
$\tan\beta$ and a model line attached to it (SPS 1a) and of a ``typical'' 
mSUGRA point with relatively high $\tan\beta$ (SPS 1b). The two-points
lie in the ``bulk'' of the cosmological region where the lightest SUSY
particle (LSP) gives rise to an acceptable dark matter density. For the 
collider phenomenology in particular the $\tau$-rich neutralino and chargino
decays are important.

\item[SPS 2: ``focus point'' scenario in mSUGRA]

\mbox{}

The benchmark point chosen for SPS~2 lies in the ``focus point'' region,
where a too large relic abundance is avoided by an enhanced annihilation
cross section of the LSP due to a sizable higgsino component.
This scenario features relatively heavy squarks and sleptons, while the
charginos and the neutralinos are fairly light and the gluino is lighter
than the squarks.

\item[SPS 3: model line into ``coannihilation region'' in mSUGRA]

\mbox{}

The model line of this scenario is directed into the ``coannihilation
region'', where a sufficiently low relic abundance can arise from
a rapid coannihilation between the LSP and the (almost mass degenerate)
next-to-lightest SUSY particle (NLSP),
which is usually the lighter $\tilde\tau$. Accordingly, an
important feature in the collider phenomenology of this scenario is the
very small slepton--neutralino mass difference.

\item[SPS 4: mSUGRA scenario with large $\tan\beta$]

\mbox{}

The large value of $\tan\be$ in this scenario has an important impact on
the phenomenology in the Higgs sector. The couplings of 
$A, H$ to $b\bar{b}$ and $\tau^+\tau^-$ as well as the 
$H^{\pm} t\bar{b}$ couplings are
significantly enhanced in this scenario, resulting in particular
in large associated production cross sections for the heavy Higgs
bosons.
 
\item[SPS 5: mSUGRA scenario with relatively light scalar top quark]

\mbox{}

This scenario is characterised by a large negative value of $A_0$, which
allows consistency of the relatively low value of $\tan\beta$ with
the constraints from the Higgs search at LEP, see
\citere{Djouadi:2001yk}.

\item[SPS 6: mSUGRA-like scenario with non-unified gaugino masses] 

\mbox{}

In this scenario, the bino mass parameter $M_1$ is larger than in the
usual mSUGRA models by a factor of $1.6$. While a bino-like neutralino is 
still the LSP, the mass difference between the lightest chargino and the
lightest two neutralinos and the sleptons is significantly reduced 
compared to the typical mSUGRA case. Neutralino, chargino and slepton decays
will feature less-energetic jets and leptons as a consequence. 
 
\item[SPS 7: GMSB scenario with $\tilde \tau$ NLSP]

\mbox{}

The NLSP in this GMSB scenario is the lighter stau, with allowed three
body decays of right-handed selectrons and smuons into it.
The decay of the NLSP into the Gravitino and the $\tau$ in this scenario
can be chosen to be prompt, delayed or quasi-stable.
 
\item[SPS 8: GMSB scenario with neutralino NLSP]

\mbox{}

The NLSP in this scenario is the lightest neutralino. The second
lightest neutralino has a significant branching ratio into $h$ when
kinematically allowed.
The decay of the NLSP into the Gravitino (and a photon or a $Z$~boson)
in this scenario can be chosen to be prompt, delayed or quasi-stable.

\item[SPS 9: AMSB scenario]

\mbox{}

This scenario features a very small neutralino--chargino mass
difference, which is typical for AMSB scenarios. Accordingly, the LSP is
a neutral wino and the NLSP a nearly degenerate charged wino. The NLSP
decays to the LSP and a soft pion with a macroscopic decay length, as
much as 10~cm.

\end{description}

As an example, below the benchmark values (i.e.\ the low-energy MSSM
parameters) are given for the benchmark point of SPS~1a. All mass
parameters are given in GeV. The value of the
top-quark mass for all SPS benchmarks is chosen to be $\mt = 175$~GeV.

All mass parameters for the benchmark point of SPS~1a 
are to be understood as defined in the 
$\overline{\rm DR}$ scheme at the scale $Q = 454.7$~GeV.\\ 
The gluino mass $\Mgl$, the Supersymmetric Higgs mass parameter $\mu$,
the mass of the $\cp$-odd Higgs boson $\MA$, the ratio of the vacuum
expectation values of the two Higgs doublets $\tan\be$, and the
electroweak gaugino mass parameters $M_1$ and $M_2$ have the following
values:
\beq
\Mgl = 595.2, \quad \mu = 352.4, \quad \MA = 393.6, \quad
\tan\be =  10, \quad M_1 = 99.1, \quad M_2 = 192.7.
\eeq
The soft SUSY-breaking parameters in the diagonal entries of the squark
and slepton mass matrices have been chosen to be the same for the first 
and second generation. They have the following values
(these parameters are approximately equal to the
sfermion masses; the off-diagonal entries have been neglected for the
first two generations; the index $i$ in $M_{\tilde qi_L}$
refers to the generation): 
\beq
M_{\tilde q1_L} = M_{\tilde q2_L} = 539.9, \quad
M_{\tilde{d}_R} = 519.5, \quad
M_{\tilde{u}_R} = 521.7, \quad
M_{\tilde{e}_L} = 196.6, \quad
M_{\tilde{e}_R} = 136.2 .
\eeq
The soft SUSY-breaking parameters in the diagonal entries of the squark and
slepton mass matrices of the third generation have the following values,
\beq
M_{\tilde q3_L} = 495.9 , \quad 
M_{\tilde{b}_R} = 516.9, \quad 
M_{\tilde{t}_R} = 424.8, \quad
M_{\tilde{\tau}_L} = 195.8, \quad
M_{\tilde{\tau}_R} = 133.6, 
\eeq
while the trilinear couplings of the third generation read
\beq
A_t = -510.0, \quad
A_b = -772.7, \quad
A_{\tau} = -254.2.
\eeq

\begin{figure}[t]
\bce
\mbox{} \hspace{3em} SPS 1a\\[.3em] 
\includegraphics[width=14cm]{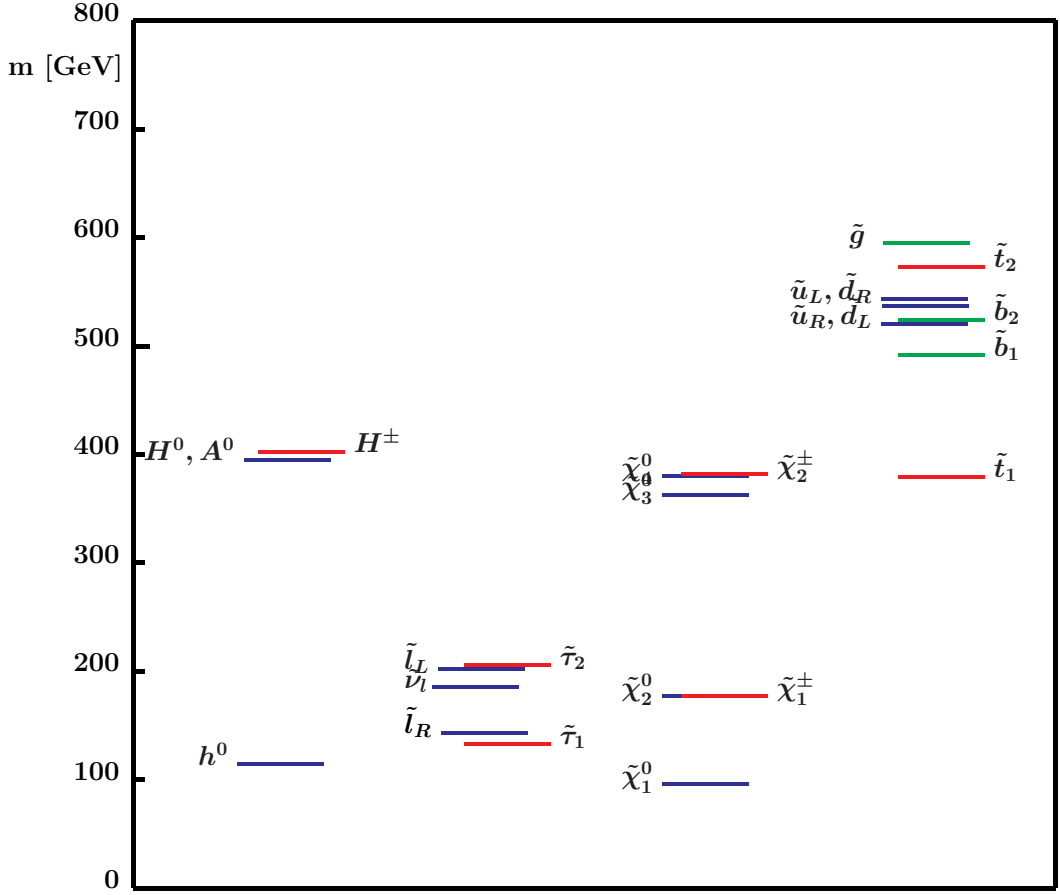}
\ece
\caption{The SUSY particle spectrum for the benchmark point
corresponding to SPS 1a (see \citeres{sps,ulinabil}).
}
\label{fig1}
\end{figure}

The corresponding SUSY particle spectrum as obtained with 
{\sl ISAJET 7.58}~\cite{isajet} is shown in \reffi{fig1}. 

The benchmark values for the other SPS can be found at 
{\tt www.ippp.dur.ac.uk/$\sim$georg/sps}, see also \citere{ulinabil}.


\subsection*{Acknowledgements}
The author thanks the members of the LHC / LC Study Group and the
authors of hep-ph/0202233 for their collaboration. I am grateful to the
organisers of SUSY02 for the invitation and the very
pleasant atmosphere at the conference. 



\end{document}